\documentclass{article}
\usepackage{amsmath}
\usepackage{amsmath}
\usepackage{amsmath}
\usepackage{amsmath}
\usepackage{amsmath}
\usepackage{amsmath}
\usepackage{amsmath}
\usepackage{amsmath}
\usepackage{amsmath}
\usepackage{amsmath}
\usepackage{amsmath}
\usepackage{amsmath}
\usepackage{amsmath}
\usepackage{amsmath}
\usepackage{amsmath}
\usepackage{amsmath}
\usepackage{amsmath}
\usepackage{amsmath}
\usepackage{amsmath}
\usepackage{amsmath}
\usepackage{amsmath}
\usepackage{amsmath}
\usepackage{amsmath}
\usepackage{amsmath}
\usepackage{amsmath}
\usepackage{amsmath}

\newtheorem{theorem}{Theorem}
\newtheorem{acknowledgement}[theorem]{Acknowledgement}

\input{tcilatex}

\begin{document}

\title{Canonical quantization of so-called non-Lagrangian systems.}
\author{D.M. Gitman$^{1}$, V.G. Kupriyanov$^{1,2}$ \\
%EndAName
\\
$^{1}$Instituto de F\'{\i}sica, Universidade de S\~{a}o Paulo,\\
Caixa Postal 66318-CEP, 05315-970 S\~{a}o Paulo, S.P., Brazil\\
$^{2}$ Physics Department, Tomsk State University, Tomsk\textit{, 634050, }%
Russia\\
E-mail: gitman@dfn.if.usp.br (D.M.Gitman), kvg@dfn.if.usp.br
(V.G.Kupriyanov).}
\date{}
\maketitle

\begin{abstract}
We present an approach to the canonical quantization of systems with
equations of motion that are historically called non-Lagrangian equations.
Our viewpoint of this problem is the following: despite the fact that a set
of differential equations cannot be directly identified with a set of
Euler-Lagrange equations, one can reformulate such a set in an equivalent
first-order form which can always be treated as the Euler-Lagrange equations
of a certain action. We construct such an action explicitly. It turns out
that in the general case the hamiltonization and canonical quantization of
such an action are non-trivial problems, since the theory involves
time-dependent constraints. We adopt the general approach of hamiltonization
and canonical quantization for such theories (Gitman, Tyutin, 1990) to the
case under consideration. There exists an ambiguity (not reduced to a total
time derivative) in associating a Lagrange function with a given set of
equations. We present a complete description of this ambiguity. The proposed
scheme is applied to the quantization of a general quadratic theory. In
addition, we consider the quantization of a damped oscillator and of a
radiating point-like charge.
\end{abstract}

\section{Introduction}

It is well-known that some physical systems like dissipative systems \cite%
{kup1}, Dirac monopole \cite{Dirac}, etc. are usually described in terms of
second-order equations of motion which cannot be directly identified with
Euler-Lagrange equations for an action principle. Following traditional
terminology, we call such equations of motion non-Lagrangian equations in
what follows. Sometimes (but not always) non-Lagrangian equations can be
reduced to Euler-Lagrange equations by multiplying by the so-called
integrating multiplier \cite{Duglas}-\cite{Sarlet}. The existence of an
action principle for a given physical system, or what is the same, the
existence of a Lagrange function for such a system, allows one to proceed
with canonical quantization schemes. This, in particular, stresses the
importance of formulating action principle for any physical system.

In the present work we discuss an approach to constructing quantum theories
that in the classical limit reproduce non-Lagrangian equations of motion for
mean values. In fact, we consider a canonical quantization of Lagrangian
theories with time-dependent constraints that are related to the
non-Lagrangian systems. To this end, on the classical level, we reduce
non-Lagrangian equations of motion to an equivalent set of first-order
differential equations. For such equations, one can always construct an
action principle, the corresponding consideration is represented in section
2 and, partially, is based on results of works \cite{Hojman}-\cite{kup}. The
hamiltonization of the constructed Lagrangian theory leads to a Hamiltonian
theory with time-dependent constraints as it is demonstrated in section 3.
Thus, we show that systems traditionally called non-Lagrangian ones are, in
fact, equivalent to some first-order Lagrangian systems, however, with
time-dependent constraints in Hamiltonian formulation. The canonical
quantization of the latter theory is not a trivial problem (it follows to
general consideration of \cite{GTbook}) and is represented in section 4. It
is known that on the classical level, there exists an ambiguity in
constructing Lagrange function (which is not reduced to a total time
derivative) for a given set of equations \cite{Hojman}-\cite{Tempesta}. We
describe completely such an ambiguity for the case under consideration. We
apply the general approach to formulate the canonical quantization in case
of theories with arbitrary linear inhomogenous equations of motion (general
quadratic theories), see section 5. Then we consider the canonical
quantization of a damped harmonic oscillator (sec. 6) and a radiating
point-like charge (sec.7).

\section{Action principle for non-Lagrangian systems}

Let a system with $n$ degrees of freedom be described by a set of $n$
non-Lagrangian second-order differential equations of motion. To construct
an action principle, we replace these equations (which is always possible)
by an equivalent set of $2n$ first-order differential equations, solvable
with respect to time derivatives. Suppose such a set has a form 
\begin{equation}
\dot{x}^{\alpha }=f^{\alpha }(t,x)\,,\;\alpha =1,..,2n\,,  \label{i1}
\end{equation}%
where $f^{\alpha }(t,x)$\thinspace are some functions of the indicated
arguments and by dots above we denote time derivatives of coordinates. Since
these equations are first-order, action $S[x]$ that yields (\ref{i1}) as
Euler--Lagrange equations, must be linear in the first time derivative $\dot{%
x}^{\alpha }$. Its general form is 
\begin{equation}
S[x]=\int dt\,L,\;\;L=J_{\alpha }\dot{x}^{\alpha }-H\,,  \label{i2}
\end{equation}%
where $J_{\alpha }=J_{\alpha }(t,x)$ and $H=H(t,x)$ are some functions of
the indicated arguments. The Euler--Lagrange equations corresponding to (\ref%
{i2}) are 
\begin{equation}
\frac{\delta S}{\delta x}=\frac{\partial L}{\partial x}-\frac{d}{dt}\frac{%
\partial L}{\partial \dot{x}}=0\Longrightarrow -\partial _{\alpha
}H-\partial _{t}J_{\alpha }+\left( \partial _{\alpha }J_{\beta }-\partial
_{\beta }J_{\alpha }\right) \dot{x}^{\beta }=0\,,  \label{i3}
\end{equation}%
where the notation are used 
\begin{equation*}
\partial _{\alpha }=\frac{\partial }{\partial x^{\alpha }}\,,\;\partial _{t}=%
\frac{\partial }{\partial t}\,.
\end{equation*}%
Denoting the combination $\left( \partial _{\alpha }J_{\beta }-\partial
_{\beta }J_{\alpha }\right) $ by $\Omega _{\alpha \beta }\,,$%
\begin{equation}
\Omega _{\alpha \beta }=\partial _{\alpha }J_{\beta }-\partial _{\beta
}J_{\alpha }=\Omega _{\alpha \beta }(t,x)=-\Omega _{\beta \alpha }(t,x)\,,
\label{i4}
\end{equation}%
we rewrite (\ref{i3}) as follows:%
\begin{equation}
\Omega _{\alpha \beta }\dot{x}^{\beta }=\partial _{\alpha }H+\partial
_{t}J_{\alpha }\,.  \label{i3a}
\end{equation}%
Equations (\ref{i3}) or (\ref{i3a}) can be identified with (\ref{i1}),
provided 
\begin{eqnarray}
&&\det \Omega _{\alpha \beta }\neq 0\,,  \label{det} \\
&&\Omega _{\alpha \beta }f^{\beta }-\partial _{t}J_{\alpha }=\partial
_{\alpha }H\,.  \label{i5}
\end{eqnarray}%
The functions $J_{\alpha }$ and $H$ can be found from conditions (\ref{i4}%
)--(\ref{i5}) if the matrix $\Omega _{\alpha \beta }$ is given. Assuming
that $J_{\alpha }$ and $H$ are smooth functions the consistency condition
for equations (\ref{i5}) imply 
\begin{equation}
\partial _{\beta }\left( \Omega _{\alpha \gamma }f^{\gamma }\right)
-\partial _{\alpha }\left( \Omega _{\beta \gamma }f^{\gamma }\right)
+\partial _{t}\Omega _{\alpha \beta }=0\Longrightarrow \partial _{t}\Omega
_{\alpha \beta }+\pounds _{f}\Omega _{\alpha \beta }=0\,,  \label{i6}
\end{equation}%
where $\pounds _{f}\Omega _{\alpha \beta }$ is the Lie derivative of $\Omega
_{\alpha \beta }$ along the vector field $f^{\gamma }.$ In addition, one can
verify that the matrix $\Omega _{\alpha \beta }$ (\ref{i4}) obeys the Jacobi
identity ($\Omega _{\alpha \beta }$ is a symplectic matrix) 
\begin{equation}
\partial _{\alpha }\Omega _{\beta \gamma }+\partial _{\beta }\Omega _{\gamma
\alpha }+\partial _{\gamma }\Omega _{\alpha \beta }=0\,.  \label{i7}
\end{equation}

Now we are going to analyze these equations. It is known that the general
solution $\Omega _{\alpha \beta }$ of equation (\ref{i6}) can be constructed
with the help of a solution of the Cauchy problem for equations (\ref{i1}).
Suppose that such a solution is known,%
\begin{equation}
x^{\alpha }=\varphi ^{\alpha }(t,x_{(0)})\,,\;x_{\left( 0\right) }^{\alpha
}=\varphi ^{\alpha }(0,x_{(0)})  \label{ii9}
\end{equation}%
be a solution of equations (\ref{i1}) for any $x_{(0)}=\left( x_{\left(
0\right) }^{\alpha }\right) ,$ and $\chi ^{\alpha }(t,x)$ be the inverse
function with respect to $\varphi ^{\alpha }(t,x_{(0)}),$ i.e.,%
\begin{equation}
x^{\alpha }=\varphi ^{\alpha }(t,x_{(0)})\Longrightarrow x_{\left( 0\right)
}^{\alpha }=\chi ^{\alpha }(t,x)\,,\;x^{\alpha }\equiv \varphi ^{\alpha
}(t,\chi ^{\alpha })\,,\;\,\partial _{\alpha }\chi ^{\gamma }|_{t=0}=\delta
_{\gamma }^{\alpha }\,.  \label{inverse}
\end{equation}%
Then 
\begin{equation}
\Omega _{\alpha \beta }(t,x)=\partial _{\alpha }\chi ^{\gamma }\,\Omega
_{\gamma \delta }^{\left( 0\right) }\left( \chi \right) \,\partial _{\beta
}\chi ^{\delta }\,,  \label{2form}
\end{equation}%
where the matrix $\Omega _{\alpha \beta }^{\left( 0\right) }$ is the initial
condition for $\Omega _{\alpha \beta }$, 
\begin{equation*}
\Omega _{\alpha \beta }(t,x)|_{t=0}=\Omega _{\alpha \beta }^{\left( 0\right)
}(x)\,.
\end{equation*}

It follows from (\ref{i7}) at $t=0$ that the matrix $\Omega _{\alpha \beta
}^{\left( 0\right) }(x)$ obeys the Jacoby identity such that the general
structure of this matrix is (we do not consider global problems which arise
from a nontrivial topology of the $x^{\alpha }$-space)%
\begin{equation}
\Omega _{\alpha \beta }^{\left( 0\right) }=\partial _{\alpha }j_{\beta
}-\partial _{\beta }j_{\alpha }\,,  \label{arbitrar}
\end{equation}%
where $j_{\alpha }\left( x\right) $ are some arbitrary functions. Then
equation (\ref{2form}) implies%
\begin{equation}
\Omega _{\alpha \beta }=\partial _{\alpha }\psi _{\beta }-\partial _{\beta
}\psi _{\alpha }\,,\;\psi _{\alpha }(t,x)=j_{\beta }\left( \chi (t,x)\right)
\partial _{\alpha }\chi ^{\beta }(t,x)\,.  \label{arbitrar1}
\end{equation}

On the other hand, relation (\ref{i4}) must hold,%
\begin{equation*}
\partial _{\alpha }\psi _{\beta }-\partial _{\beta }\psi _{\alpha }=\partial
_{\alpha }J_{\beta }-\partial _{\beta }J_{\alpha }\,,
\end{equation*}%
which implies that%
\begin{equation}
J_{\alpha }(t,x)=\psi _{\alpha }+\partial _{\alpha }\varphi =j_{\beta
}\left( \chi (t,x)\right) \partial _{\alpha }\chi ^{\beta }(t,x)+\partial
_{\alpha }\varphi (t,x)\,,  \label{ji}
\end{equation}%
where $\varphi (t,x)$ is an arbitrary function. One can represent another
form for $J_{\alpha }(t,x),$ in which the ambiguity related to the arbitrary
functions $j_{\beta }\left( x\right) $ is incorporated in the matrix $\Omega
_{\alpha \beta }^{\left( 0\right) }.$ To this end, we remind that the
general solution for $J_{\alpha }(t,x)$ of the equation (\ref{i4}) provided
that $\Omega _{\alpha \beta }$ is a given antisymmetric matrix that obeys
the Jacobi identity, is given by 
\begin{equation}
J_{\alpha }(t,x)=\int_{0}^{1}x^{\beta }\Omega _{\beta \alpha
}(t,sx)\,sds+\partial _{\alpha }\varphi (t,x)\,,  \label{ji2}
\end{equation}%
where $\varphi (x)$ is an arbitrary function. Substituting (\ref{2form})
into (\ref{ji2}), we obtain 
\begin{equation}
J_{\alpha }(t,y)=\int_{0}^{1}y^{\beta }\left. \left[ \partial _{\alpha }\chi
^{\gamma }\,\Omega _{\gamma \delta }^{\left( 0\right) }\left( \chi \right)
\,\partial _{\beta }\chi ^{\delta }\right] \right| _{x=sy}\,\,sds+\partial
_{\alpha }\varphi (t,y)\,.  \label{ji3}
\end{equation}%
Equations (\ref{ji}) or (\ref{ij3}) desribe all the ambiguity (arbitrary
functions $j_{\beta }\left( x\right) $ and $\varphi (t,x),$ or arbitrary
symplectic matrix $\Omega _{\gamma \delta }^{\left( 0\right) }$ and
arbitrary function $\varphi (t,x)$) in constructing the term $J_{\alpha
}(t,x)$ of the Lagrange function (\ref{i2}).

One can also see that choosing the matrix $\Omega _{\alpha \beta }^{\left(
0\right) }(x)$ to be nonsingular, we guarantee the nonsingularity (condition
(\ref{det})) for the matrix $\Omega _{\alpha \beta }(t,x)$ since components
of the latter are given by a change of variables (\ref{2form}).

To restore the term $H$ in the Lagrange function (\ref{i2}), we need to
solve the equation (\ref{i5}) with respect to $H.$ To this end, we remind
that the general solution of the equation $\partial _{i}f=g_{i}$, provided a
vector $g_{i}$ is a gradient, is given by 
\begin{equation*}
f(x)=\int_{0}^{1}ds\,\,x^{i}g_{i}(sx)\,+c\,,
\end{equation*}%
where $c$ is a constant. Taking the above into account, we obtain for $H$
the following representation: 
\begin{equation}
H(t,x)=\int_{0}^{1}ds\,x^{\beta }\left[ \Omega _{\beta \alpha
}(t,sx)f^{\alpha }(t,sx)-\partial _{t}J_{\beta }(t,sx)\right] +c(t)\,,
\label{i9}
\end{equation}%
where $c(t)$ is an arbitrary function of time, and $\Omega _{\beta \alpha }$
and $J_{\beta }$ are given by (\ref{2form}) and (\ref{ji3}) respectively.
All the arbitrarness in constructing $H$ is thus due to arbitrary symplectic
matrix $\Omega _{\gamma \delta }^{\left( 0\right) }$, arbitrary functions $%
\varphi (t,x)$ entering into $\Omega _{\beta \alpha }$ and $J_{\beta }$ and
due $c(t).$

We see that there exist a family of actions (\ref{i2}) which lead to the
same equations of motion (\ref{i1}). It is easy to see that actions with the
same $\Omega _{\gamma \delta }^{\left( 0\right) }$ but different functions $%
\varphi (t,x)$ and $c(t)$ differs by a total time derivative (we call such a
difference trivial). A difference in Lagrange functions related to different
choice of symplectic matrices $\Omega _{\alpha \beta }^{\left( 0\right) }$
is not trivial. The coresponding Lagrangians are known as $s$-equivalent
Lagrangians. In spite of the fact that actions with nontrivial difference
lead to the same equations of motion, they lead in general to different
Hamiltonian formulations and to different quantum theories in course of the
quantization. However, any quantum theory that is obtained by the developed
below quantization procedure obeys the correspondence principle, i.e., in
the classical limit, equations of motion for mean values coincide with (\ref%
{i1}). Equations of motion (\ref{i1}) do no contain any additional
information that can be used in choosing a ''right'' quantum theory. Only
physical considerations or a comparison with experiment may be used for this
aim. Below, we are going to consider hamiltonization and subsequent
quantization of the action (\ref{i2}) with the following choice of the
symplectic matrix $\Omega _{\gamma \delta }^{\left( 0\right) }$%
\begin{equation}
\Omega _{\alpha \beta }^{\left( 0\right) }=\left( 
\begin{array}{cc}
\mathbf{0} & -\mathbf{I} \\ 
\mathbf{I} & \mathbf{0}%
\end{array}%
\right) ,  \label{ic}
\end{equation}%
where $\mathbf{I}$ is an $n\times n$ unit matrix, and $\mathbf{0}$ denotes
an $n\times n$ zero matrix. This choice of the action leads to the canonical
commutation relations for original variables on the quantum level.
Hamiltonization and quantization of the action (\ref{i2}) with different
choices of the symplectic matrix $\Omega _{\gamma \delta }^{\left( 0\right)
} $ can be fulfilled in the same manner, but technically look more clumsily.

Note that (\ref{ic}) implies that there exist only two possibilities for the
matrix $\Omega $\ in (\ref{2form}). Namely, it is either a canonical
symplectic matrix, which is possible only if the initial equations (\ref{i1}%
) are canonical Hamiltonian equations (or, equivalently, Lagrangian
equations of motion for the first-order action), or it must depend on time,
which is the case of non-Lagrangian equations.

The first-order action (\ref{i2}) can be regarded as a Lagrangian action, or
as a Hamiltonian action with a noncanonical Poisson bracket. An equivalent
second-order Lagrangian formulation is always possible; however, it may
include additional variables \cite{GLNT}.

One ought to say that it is always possible to construct a Lagrangian action
for non-Lagrangian second-order equations in an extended configuration space
following a simple idea first proposed by Bateman \cite{Bateman}. Such a
Lagrangian has a form of a sum of initial equations of motion being
multiplied by the corresponding Lagrangian multipliers, new variables.
Euler-Lagrange equations for such an action contain besides the initial
equations some new equations of motion for the Lagrange multipliers. In such
an approach one has to think how to interprete the new variables already on
the classical level. Additional difficulties (indefinite metric) can appear
in course of the quantization.

As an example, we consider a theory with equations of motion of the form%
\footnote{%
Here we use matrix notation, $x=\left( x^{\alpha }\right) ,\;A(t)=\left(
A(t)_{\beta }^{\alpha }\right) ,\;j(t)=\left( j(t)^{\alpha }\right)
,\;\alpha ,\beta =1,..,2n$.
\par
{}} 
\begin{equation}
\dot{x}=A(t)x+j(t)\,.  \label{i10}
\end{equation}
We call such a theory the general quadratic theory. Let us apply the above
consideration to construct the action principle for such a theory.

Solution of the Cauchy problem for the equations (\ref{i10}) reads 
\begin{equation}
x(t)=\Gamma (t)x_{(0)}+\gamma (t)\,,  \label{i11}
\end{equation}%
where the matrix $\Gamma (t)$ is the fundamental solution of (\ref{i10}),
i.e., 
\begin{equation}
\dot{\Gamma}=A\Gamma \,,\;\Gamma (0)=1\,,  \label{i12}
\end{equation}%
and $\gamma (t)$ is a partial solution of (\ref{i10}).\ Then following (\ref%
{2form}), we construct the matrix $\Omega $, 
\begin{equation}
\Omega =\Lambda ^{T}\Omega ^{\left( 0\right) }\Lambda \,,\;\;\Lambda =\Gamma
^{-1}\,.  \label{i13}
\end{equation}%
and find the functions $J$ and $H$ according to (\ref{ji3}) and (\ref{i9}), 
\begin{equation}
J=\frac{1}{2}x\Omega \,,\;H=\frac{1}{2}xBx-Cx\,,  \label{i14}
\end{equation}%
where 
\begin{equation}
B=\frac{1}{2}\left( \Omega A-A^{T}\Omega \right) \,,\;C=\Omega j\,.
\label{i15}
\end{equation}%
Thus, the action functional for the general quadratic theory is 
\begin{equation}
S[x]=\frac{1}{2}\int dt\left( x\Omega \dot{x}-xBx-2Cx\right) \,.  \label{i16}
\end{equation}%
Another approach to constructing the action functional for the general
quadratic theory was proposed in \cite{kup1}.

Note that Darboux coordinates $x_{0}$ can be written via a matrix $\Lambda $
as follows: 
\begin{equation}
x\rightarrow x_{0}=R^{-1}(t)\Lambda (t)x\,.  \label{i17}
\end{equation}
Here, $R(t)$ is an arbitrary matrix of a linear (generally time-dependent)
canonical transformation: 
\begin{equation*}
R^{T}(t)\Omega ^{\left( 0\right) }R(t)=\Omega ^{\left( 0\right) }\,.
\end{equation*}
In terms of the coordinates $x_{0}$, action (\ref{i16}) takes the form 
\begin{equation}
S[x]=\frac{1}{2}\int dt\left( x_{0}\Omega ^{\left( 0\right) }\dot{x}%
_{0}+x_{0}R^{T}\Omega ^{\left( 0\right) }\dot{R}x_{0}-2C\Gamma Rx_{0}\right)
\,.  \label{i18}
\end{equation}

The Darboux coordinates (\ref{i17}) can be divided into coordinates and
corresponding momenta. The Euler--Lagrange equations for action (\ref{i18})
have the form of canonical Hamilton equations with the Hamiltonian 
\begin{equation}
H_{0}=-\frac{1}{2}x_{0}R^{T}\Omega ^{\left( 0\right) }\dot{R}x_{0}+C\Gamma
Rx_{0}\,.  \label{i19}
\end{equation}

Note that the choice $R=\mathrm{const}$ yields a trivial Hamiltonian, which
is consistent with the fact that in this case $x_{0}$ are the initial data
without dynamics.

\section{Hamiltonian formulation}

We are now going to consider action (\ref{i2}) as a Lagrangian action with
the Lagrange function 
\begin{equation}
L=J_{\alpha }\left( t,x\right) \dot{x}^{\alpha }-H\left( t,x\right) \,
\label{3.1}
\end{equation}
and construct a corresponding Hamiltonian formulation. To this end, we
follow the general\footnote{%
Note that some of $J_{\alpha }$ can be equal to zero, for instance, if one
deals with a canonical Hamiltonian action. In this case, one obtains the
constraints $\Phi _{\alpha }=\pi _{\alpha }=0$. Another way to examine this
case is to use the method of hamiltonization for theories with degenerate
coordinates \cite{GT1}.} scheme of \cite{GTbook}. We first construct the
action $S^{v}[x,\pi ,v],$ which, in this case, has the form 
\begin{equation}
S^{v}[x,\pi ,v]=\int \left[ J_{\alpha }\left( t,x\right) v^{\alpha }-H\left(
t,x\right) +\pi _{\alpha }\left( \dot{x}^{\alpha }-v^{\alpha }\right) \right]
dt\,,  \label{3.2}
\end{equation}
and depends on the momenta $\pi _{\alpha }$ conjugate to the coordinates $%
x^{\alpha }$, as well as on the velocities $v^{\alpha }.$ The equations 
\begin{equation}
\frac{\delta S^{v}}{\delta v^{\alpha }}=\Phi _{\alpha }\left( t,x,\pi
\right) =\pi _{\alpha }-J_{\alpha }(t,x)=0  \label{3.3}
\end{equation}
do not allow one to express the velocities via $x$ and $\pi $, which implies
the appearance of primary constraints $\Phi _{\alpha }\left( t,x,\pi \right) 
$, and the velocities $v^{\alpha }$ become Lagrangian multipliers to these
constraints, so that action (\ref{3.2}) becomes a Hamiltonian action of a
theory with the primary constraints (\ref{3.3}), 
\begin{equation}
S_{\mathrm{H}}=\int dt\{\pi _{\alpha }\dot{x}^{\alpha }-H^{\left( 1\right)
}\}\,,\;H^{(1)}=H(t,x)+\lambda ^{\alpha }\Phi _{\alpha }\left( t,x,\pi
\right) \,,  \label{3.4}
\end{equation}
with the equations of motion 
\begin{equation}
\dot{\eta}=\left\{ \eta ,H^{(1)}\right\} \,,\;\Phi =0\,,  \label{3.4a}
\end{equation}
where $\eta =(x,\pi )$.

The primary constraints are second-class ones. Indeed, we have, in virtue of
(\ref{det}), 
\begin{equation}
\{\Phi _{\alpha },\Phi _{\beta }\}=\Omega _{\alpha \beta
}(t,x)\Longrightarrow \det \{\Phi _{\alpha },\Phi _{\beta }\}\neq 0\,.
\label{3.5}
\end{equation}
Thus, secondary constraints do not appear, and all $\lambda $-s are
determined from the consistency conditions for the primary constraints: 
\begin{eqnarray}
&&\dot{\Phi}_{\alpha }=\partial _{t}\Phi _{\alpha }+\{\Phi _{\alpha
},H^{(1)}\}=0\Longrightarrow -\partial _{t}J_{\alpha }-\partial _{\alpha
}H+\lambda ^{\beta }\{\Phi _{\alpha },\Phi _{\beta }\}=0\Longrightarrow 
\notag \\
&&\,\lambda ^{\beta }=\omega ^{\beta \alpha }\left( \partial _{t}J_{\alpha
}+\partial _{\alpha }H\right) \,,\;\omega ^{\beta \alpha }=\Omega _{\beta
\alpha }^{-1}\,.  \label{3.6}
\end{eqnarray}
Using the Lagrange multipliers (\ref{3.6}) in equations (\ref{3.4a}), we can
write these equations in the form 
\begin{equation}
\dot{\eta}=\{\eta ,H\}_{D(\Phi )}+\{\eta ,\Phi _{\alpha }\}\omega ^{\alpha
\beta }\partial _{t}J_{\beta }\,,\;\Phi =0\,,  \label{3.7}
\end{equation}
where $\{\cdots ,\cdots \}_{D(\Phi )}$ are the Dirac brackets with respect
to the second-class constraints $\Phi .$ For the canonical variables the
Dirac brackets are 
\begin{eqnarray}
&\{x^{\alpha },x^{\beta }\}_{D(\Phi )}=&\omega ^{\alpha \beta }\,,  \notag \\
&\{\pi _{\alpha },\pi _{\beta }\}_{D(\Phi )}=&\partial _{\alpha }J_{\rho
}\omega ^{\rho \gamma }\partial _{\beta }J_{\gamma }\,,  \notag \\
&\{x^{\alpha },\pi _{\beta }\}_{D(\Phi )}=&\delta _{\beta }^{\alpha }+\omega
^{\alpha \gamma }\partial _{\beta }J_{\gamma }\,.  \label{a34}
\end{eqnarray}
Formally introducing a momentum $\epsilon $ conjugate to the time $t$, and
defining the Poisson brackets in an extended space of the canonical
variables $(x,\pi ;t,\epsilon )=(\eta ;t,\epsilon ),$ see \cite{GTbook}, we
can rewrite (\ref{3.7}) as follows: 
\begin{equation}
\dot{\eta}=\{\eta ,H+\epsilon \}_{D(\Phi )}\,,\;\Phi =0\,.  \label{3.8}
\end{equation}

Equations (\ref{3.8}) present a Hamiltonian formulation of non-Lagrangian
systems with first-order equations of motion (\ref{i1}). We note that the
Hamiltonian constraints in this formulation are second-class ones and depend
on time explicitly. The\ canonical quantization of theories with
time-dependent second-class constraints can be carried out along the lines
of \cite{GTbook}. Below, we present the details of such a quantization, and
then adopt it to the system under consideration.

\section{Canonical quantization}

For a Hamiltonian theory with time-dependent second-class constraints, the
quantization procedure in the ``Schr\"{o}dinger''\ picture is realized as
follows. The phase-space variables $\eta $ of a theory with time-dependent
second-class constraints $\Phi _{l}(\eta ,t)$\ are assigned operators $\hat{%
\eta}\left( t\right) $ subject to the equal-time commutation relations and
the constraints equations 
\begin{equation}
\lbrack \hat{\eta}^{A}\left( t\right) ,\hat{\eta}^{B}\left( t\right)
]=i\{\eta ^{A},\eta ^{B}\}_{D(\Phi )}|_{\eta =\hat{\eta}}\,,\;\Phi _{l}(\hat{%
\eta}\left( t\right) ,t)=0\,.  \label{e2}
\end{equation}
Their time evolution is postulated as (we neglect the problem of operator
ordering \cite{Berezin}) 
\begin{equation}
\frac{d}{dt}\hat{\eta}\left( t\right) =\{\eta ,\epsilon \}_{D(\Phi )}|_{\eta
=\hat{\eta}}=-\{\eta ,\Phi _{l}\}\{\Phi ,\Phi \}_{ll^{\prime }}^{-1}\partial
_{t}\Phi _{l^{\prime }}|_{\eta =\hat{\eta}}\;.  \label{e4}
\end{equation}
To each physical quantity $F$, given in the Hamiltonian formulation by a
function $F(t,\eta )$, we assign a ``Schr\"{o}dinger''\ operator $\hat{F}%
\left( t\right) ,$ by the rule $\hat{F}\left( t\right) =F(t,\hat{\eta}\left(
t\right) )$. For arbitrary ``Schr\"{o}dinger''\ operators $\hat{F}\left(
t\right) $ and $\hat{G}\left( t\right) ,$ the relation 
\begin{equation}
\lbrack \hat{F}\left( t\right) ,\hat{G}\left( t\right) ]=i\{F,G\}_{D(\Phi
)}|_{\eta =\hat{\eta}}  \label{e7}
\end{equation}
holds as a consequence of (\ref{e2}). The quantum states of the system are
described by vectors $\Psi $ of a Hilbert space with a scalar product $%
\left( \Psi ,\Psi ^{\prime }\right) $. Their time evolution is determined by
the Schr\"{o}dinger equation 
\begin{equation}
i\frac{\partial \Psi \left( t\right) }{\partial t}=\hat{H}\Psi \left(
t\right) \,,  \label{e5}
\end{equation}
where the quantum Hamiltonian $\hat{H}$ is constructed according to the
classical function $H(t,\eta )$ as $\hat{H}\left( t\right) =H(t,\hat{\eta}%
\left( t\right) ).$ The mean values $\langle F\rangle _{t}$ of a physical
quantity $F$ are determined as the mean values of a corresponding ``Schr\"{o}%
dinger''\ operator $\hat{F}\left( t\right) =F(t,\hat{\eta}\left( t\right) )$
with respect to state vectors $\Psi \left( t\right) ,$%
\begin{equation}
\langle F\rangle _{t}=\left( \Psi \left( t\right) ,\hat{F}\left( t\right)
\Psi \left( t\right) \right) \,.  \label{e5a}
\end{equation}
Provided that $\hat{H}$ is a self-adjoint operator, the time evolution of
state vectors $\Psi \left( t\right) $ is unitary, 
\begin{equation}
\Psi \left( t\right) =U\left( t\right) \Psi \left( 0\right) \,,\;U^{+}\left(
t\right) =U^{-1}\left( t\right) \,,  \label{e6}
\end{equation}
where $U\left( t\right) $ is an evolution operator.

In the Heisenberg picture, where state vectors are \textquotedblleft
frozen\textquotedblright\ and the time evolution is governed by the
Heisenberg operators $\check{\eta}\left( t\right) =U^{-1}\left( t\right) 
\hat{\eta}\left( t\right) U\left( t\right) $, one can see \cite{GTbook} that 
\begin{eqnarray}
&&\frac{d}{dt}\check{\eta}=\{\eta ,H(t,\eta )+\epsilon \}_{D(\Phi )}|_{\eta =%
\check{\eta}}\,,  \notag \\
&&\,[\check{\eta}^{A}\left( t\right) ,\check{\eta}^{B}\left( t\right)
]=i\{\eta ^{A},\eta ^{B}\}_{D(\Phi )}|_{\eta =\check{\eta}}\,,\;\;\Phi (%
\check{\eta}\left( t\right) ,t)=0\,,  \label{e8}
\end{eqnarray}%
while for Heisenberg operators $\check{F}\left( t\right) =U^{-1}\left(
t\right) \hat{F}\left( t\right) U\left( t\right) =F(t,\check{\eta}\left(
t\right) ),$ we have 
\begin{equation}
\frac{d}{dt}\check{F}\left( t\right) =\{F(t,\eta ),H(t,\eta )+\epsilon
\}_{D(\Phi )}|_{\eta =\check{\eta}}\,,  \label{e11}
\end{equation}%
or 
\begin{equation}
\frac{d}{dt}\check{F}\left( t\right) =-i\left[ \check{F}\left( t\right) ,%
\check{H}\left( t\right) \right] +\{F(t,\eta ),\epsilon \}_{D(\Phi )}|_{\eta
=\check{\eta}}\,.  \label{e12}
\end{equation}%
The mean values $\langle F\rangle _{t}$ in the Heisenber picture in
according to (\ref{e5a}) and (\ref{e6}) are determined as 
\begin{equation}
\langle F\rangle _{t}=\left( \Psi \left( 0\right) ,\check{F}\left( t\right)
\Psi \left( 0\right) \right) ~.  \label{e10}
\end{equation}%
\ The above quantization provides the fulfilment of the correspondence
principle because quantum equations (\ref{e8}) has the same form as the
classical one (\ref{3.8}).

Note that the time-dependence of the Heisenberg operators in the theories
under consideration is not unitary in the general case. In other words,
there exists no such (``Hamiltonian'') operator whose commutator with a
physical quantity can produce its total time derivative. This is explained
by the existence of two factors which determine the time evolution of a
Heisenberg operator. The first one is the unitary evolution of a state
vector in the ``Schr\"{o}dinger'' picture, while the second one is the time
variation of a ``Schr\"{o}dinger'' operators $\hat{\eta}$, which in general
has a non-unitary character. The existence of these two factors is related
to the division of the right-hand side of (\ref{e12}) into two summands.
Physically, this is explained by the fact that dynamics develops on a
surface which changes with time -- in the general case, in a nonunitary way.

Below, we apply the above quantization scheme to the system under
consideration. Taking into account the Dirac brackets (\ref{a34}), we can
write the equal-time commutation relations (\ref{e2}) for phase-space
operators as 
\begin{eqnarray}
&&\left[ \hat{x}^{\alpha },\hat{x}^{\beta }\right] =i\omega ^{\alpha \beta
}|_{x=\hat{x}}\,,  \notag \\
&&\left[ \hat{\pi}_{\alpha },\hat{\pi}_{\beta }\right] =i\left. \partial
_{\alpha }J_{\rho }\omega ^{\rho \gamma }\partial _{\beta }J_{\gamma
}\right| _{x=\hat{x}}\,,  \label{a35} \\
&&\left[ \hat{x}^{\alpha },\hat{\pi}_{\beta }\right] =i\delta _{\beta
}^{\alpha }+i\left. \omega ^{\alpha \gamma }\partial _{\beta }J_{\gamma
}\right| _{x=\hat{x}}\,.  \notag
\end{eqnarray}
In this case, the classical Hamiltonian $H$ does not depend on the momenta $%
\pi _{\alpha }$, and therefore in order to determine the quantum Hamiltonian 
$\hat{H}$, we need to know only the time dependence of the operators $\hat{x}%
^{\alpha }$. From (\ref{e4}) it follows that 
\begin{equation}
\frac{d}{dt}\hat{x}^{\alpha }=\left. \omega ^{\alpha \beta }(t,x)\partial
_{t}J_{\beta }(t,x)\right| _{x=\hat{x}}\,.  \label{a41}
\end{equation}

\section{Quantization of general quadratic theory}

The quantum-mechanical description of quadratic systems is a widely
discussed physical problem which has a number of important applications
(see, e.g., \cite{Lewis}--\cite{Kiu} and references therein). Almost all of
these works deal with the case of ``Hamiltonian''\ quadratic systems, i.e.,
systems described by canonical Hamiltonian equations of motion. On the other
hand, we consider a general quadratic system, i.e., a system described by
arbitrary linear inhomogeneous equations of motion (\ref{i10}). In this case
conditions (\ref{a35}), (\ref{a41}) become 
\begin{eqnarray}
&&\left[ \hat{x}^{\alpha },\hat{x}^{\beta }\right] =i\omega ^{\alpha \beta
}(t)\,,  \label{a42} \\
&&\frac{d}{dt}\hat{x}^{\alpha }=-\frac{1}{2}\omega ^{\alpha \beta }(t)\dot{%
\Omega}_{\beta \gamma }(t)\hat{x}^{\gamma }.  \label{a43}
\end{eqnarray}
The time-dependence of the operators $\hat{x}$ can be easily found: 
\begin{equation}
\hat{x}^{\alpha }(t)=\Phi ^{\alpha \beta }(t)\hat{x}_{0}^{\beta }\,.
\label{a44}
\end{equation}
Here, the matrix $\Phi $ obeys the equation 
\begin{equation}
\dot{\Phi}=-\frac{1}{2}\omega \dot{\Omega}\Phi \,,\;\;\Phi (0)=E\,,
\label{a45}
\end{equation}
and the operators $\hat{x}_{0}$ obey the following commutation relations: 
\begin{equation}
\left[ \hat{x}_{0}^{\alpha },\hat{x}_{0}^{\beta }\right] =i\left( \Omega
_{\alpha \beta }^{\left( 0\right) }\right) ^{-1}=i\left( 
\begin{array}{cc}
\mathbf{0} & \mathbf{I} \\ 
-\mathbf{I} & \mathbf{0}%
\end{array}
\right) ,  \label{cc}
\end{equation}
see (\ref{ic}). In what follows, it is useful to divide the operators $\hat{x%
}_{0}^{\alpha }$ into the operators of coordinates proper\ and corresponding
momenta $\hat{x}_{0}^{\alpha }=(\hat{q}^{i},\hat{p}_{i})$, $\alpha =1,..,2n,$
$i=1,..,n$. The operators $\hat{q}$ and $\hat{p}$ obey the canonical
commutation relations 
\begin{equation}
\left[ \hat{q}^{i},\hat{p}_{j}\right] =i\delta _{j}^{i},\;\left[ \hat{q}^{i},%
\hat{q}^{j}\right] =\left[ \hat{p}_{i},\hat{p}_{j}\right] =0.  \label{cc1}
\end{equation}
The quantum Hamiltonian in eq. (\ref{e5}) takes the form 
\begin{equation}
\hat{H}=\frac{1}{2}\hat{x}_{0}\Phi ^{T}B\Phi \hat{x}_{0}-C\Phi \hat{x}_{0}\,,
\label{43}
\end{equation}
where the matrix $B$ is determined by (\ref{i15}).

The above quantization is equivalent to quantization in Darboux coordinates,
and the transformation$\;x\rightarrow \Phi (t)x_{0}$ provides, by itself, a
passage to the Darboux coordinates $x_{0}$, because (\ref{a45}) implies 
\begin{equation}
\Phi ^{t}\Omega \Phi =\Omega _{0}\,.  \label{a47}
\end{equation}
Namely, in the coordinates $x_{0}$ the Poisson bracket is canonical.
Therefore, $\Phi =\Gamma (t)R(t),$ where $\Gamma (t)$ is a fundamental
solution of system (\ref{i10}). However, in contrast to the classical
theory, now the matrix $R(t)$ is fixed, it must obey the conditions 
\begin{equation}
\dot{R}=\Omega _{0}\Gamma ^{t}B\Gamma R\,,\;\;R(0)=E\,.  \label{a48}
\end{equation}
Thus, using (\ref{i19}) one can also rewrite the Hamiltonian in (\ref{43})
as follows: 
\begin{equation}
\hat{H}=-\frac{1}{2}\hat{x}_{0}R^{T}\Omega ^{\left( 0\right) }\dot{R}\hat{x}%
_{0}+C\Gamma R\hat{x}_{0}\;.  \label{a49}
\end{equation}
It is remarkable that if the matrix $A$ that determines the set of equations
(\ref{i10}) is constant, the matrix that determines the quadratic part of
the Hamiltonian in\ (\ref{43}) is constant as well, and equals to 
\begin{equation}
\Phi ^{T}B\Phi =B(0)=\frac{1}{2}\left( \Omega ^{\left( 0\right)
}A-A^{T}\Omega ^{\left( 0\right) }\right) .  \label{a50}
\end{equation}
This fact is easy to observe because the time derivative of this matrix, in
view of (\ref{a45}), (\ref{i13}) and (\ref{i15}), is equal to zero: 
\begin{equation*}
\frac{d}{dt}\left( \Phi ^{T}B\Phi \right) =0.
\end{equation*}
Thus, in this case, as distinct from the general case, the matrix $\Phi $
can be determined from the set of algebraic equations (\ref{a47}) and (\ref%
{a50}).

Note that if we start from a canonical Hamiltonian system the above
quantization coincides with the usual canonical quantization, because in
this case equation (\ref{a43}) becomes $d\hat{x}/dt=0$, i.e., $\hat{x}(t)=$ $%
\hat{x}_{0}.$

In the Heisenberg picture, equations (\ref{e8}) for the operators $\check{x}$
take the form 
\begin{eqnarray}
&&\frac{d}{dt}\check{x}=A(t)\check{x}+j(t),  \label{h1} \\
&&\left[ \check{x}^{\alpha },\check{x}^{\beta }\right] =i\omega ^{\alpha
\beta }(t)\,\,.  \label{h2}
\end{eqnarray}
Equations (\ref{h1}) coincide (the correspondence principle) with the
classical equations of motion (\ref{i10}); however, the commutation
relations (\ref{h2}) differ from the canonical ones. So, evolution of
operators $\check{x}$ can be written as 
\begin{equation}
\check{x}\left( t\right) =\Gamma (t)\check{x}_{0}+\gamma (t)\,,  \label{h3}
\end{equation}
where operators $\check{x}_{0}$ as well as $\hat{x}_{0}$ obey the canonical
commutation relations 
\begin{equation}
\left[ \check{x}_{0}^{\alpha },\check{x}_{0}^{\beta }\right] =i\left( \Omega
_{\alpha \beta }^{\left( 0\right) }\right) ^{-1}=i\left( 
\begin{array}{cc}
\mathbf{0} & \mathbf{I} \\ 
-\mathbf{I} & \mathbf{0}%
\end{array}
\right) .  \label{h4}
\end{equation}
Thus, mean values $\langle F\rangle _{t}$ of a physical quantity $F$
according to (\ref{e10}) are determined as mean values of the corresponding\
operator $\check{F}\left( t\right) =F(t,\check{x}\left( t\right) )$ with
respect to initial states vectors $\Psi \left( 0\right) ,$ i.e. 
\begin{equation}
\langle F\rangle _{t}=\left( \Psi \left( 0\right) ,F\left( t,\Gamma (t)%
\check{x}_{0}+\gamma (t)\right) \Psi \left( 0\right) \right) \,.  \label{h5}
\end{equation}
We see that the quantum evolution of physical quantities in general
quadratic systems is completely determined by the classical one.

\section{Quantization of damped harmonic oscillator}

The above formulated quantization of non-Lagrangian theories and, in
particular, of general quadratic theories can be immediately applied to
quantizing a damped harmonic oscillator. The latter problem attracts
attention for already more then 50 years , there exist different approaches
to its solution, no one of them seems to be a final version which does not
contain weak points, see e.g. \cite{Kanai}-\cite{Darek}, \cite{kup1}.

The classical equation of motion for a damped harmonic oscillator is
non-Lagrangian, it has the form 
\begin{equation}
\ddot{r}+2\alpha \dot{r}+\omega ^{2}r=0\,,  \label{dos}
\end{equation}
where $\omega $ is the angular frequency and $\alpha \geq 0$ is a friction
coefficient. Introducing an auxiliary variable $y=\dot{r},$ we reduce (\ref%
{dos}) to the following equivalent pair of first-order equations: 
\begin{equation}
\dot{r}=y\,,\;\dot{y}=-\omega ^{2}r-2{\alpha }y\,.  \label{i20}
\end{equation}
Following the way proposed in sec. 2, we construct an action $S$ that
implies (\ref{i20}) as Euler-Lagrange equations, 
\begin{equation}
S=\frac{1}{2}\int dt\left[ y\dot{r}-r\dot{y}-(y^{2}+2{\alpha ry+}\omega
^{2}r^{2})\right] e^{2\alpha t}\,.  \label{i21}
\end{equation}
Note that equation (\ref{dos}) can be represented as 
\begin{equation*}
\frac{d}{dt}\left( e^{2\alpha t}\dot{r}\right) +e^{2\alpha t}\omega
^{2}r=0\,,
\end{equation*}
i.e., as a Lagrangian equation of motion with time-dependent mass and
frequency. In this case, the mass $e^{2\alpha t}$\ is nothing else but an
integrating multiplier for equation (\ref{dos}); however, as was already
mentioned, an integrating multiplier does not always exist \cite{Duglas,kup}.

Then we proceed with the canonical quantization described in the previous
section. Equal-time commutations relations (\ref{a42}) and equations (\ref%
{a43}) determining time evolution of ``Schr\"{o}dinger'' operators $\hat{r}$
and $\hat{y}$ are 
\begin{eqnarray}
\left[ \hat{r},\hat{y}\right] &=&ie^{-2\alpha t},\;\left[ \hat{r},\hat{r}%
\right] =\left[ \hat{y},\hat{y}\right] =0,  \label{f2} \\
\frac{d}{dt}\hat{r} &=&-\alpha \hat{r},\,\ \ \frac{d}{dt}\hat{y}=-\alpha 
\hat{y}\,.  \label{f1}
\end{eqnarray}
A solution of these equations has the form 
\begin{equation}
\hat{r}=e^{-\alpha t}\hat{q}\,,\ \ \hat{y}=e^{-\alpha t}\hat{p}\,,
\label{f3}
\end{equation}
where operators $\hat{q}$ and $\hat{p}$ obey canonical commutations
relations 
\begin{equation*}
\left[ \hat{q},\hat{p}\right] =i\,,\;\left[ \hat{q},\hat{q}\right] =\left[ 
\hat{p},\hat{p}\right] =0\,.
\end{equation*}
According to (\ref{43}), the corresponding quantum Hamiltonian reads 
\begin{equation*}
\hat{H}=\frac{1}{2}\left[ \hat{p}^{2}+{\alpha }\left( {\hat{q}\hat{p}+\hat{p}%
\hat{q}}\right) {+}\omega ^{2}\hat{q}^{2}\right] \,.
\end{equation*}
It can be modified to the form 
\begin{equation}
\hat{H}=\frac{1}{2}\left[ {\hat{P}}^{2}+\left( \omega ^{2}-{\alpha }%
^{2}\right) \hat{Q}^{2}\right]  \label{i22}
\end{equation}
by the help of the canonical transformation $(\hat{p},{\hat{q})}\rightarrow (%
{\hat{P}},\hat{Q})$, where ${\hat{P}=\hat{p}+\alpha \hat{q}}${, and\ }${\hat{%
Q}=\hat{q}}$. The corresponding generating function is $W=qP-\alpha q^{2}/2.$

As usual we define the classical energy of the system by 
\begin{equation*}
E=\frac{1}{2}\left( \dot{r}^{2}+\omega ^{2}r^{2}\right) =\frac{1}{2}\left(
y^{2}+\omega ^{2}r^{2}\right) \,.
\end{equation*}
One can easy see the energy depends of time as follows: $E=E_{0}e^{-2\alpha
t}$. Using (\ref{f3}), we obtain an expression for the operator $\hat{E}$
that corresponds to the classical quantity $E,$%
\begin{equation}
\hat{E}=\frac{1}{2}e^{-2\alpha t}\left[ {\hat{P}}^{2}-{\alpha (\hat{P}\hat{Q}%
+\hat{Q}\hat{P})+(}\omega ^{2}+\alpha ^{2}){\hat{Q}}^{2}\right] \,.
\label{i25}
\end{equation}

Let us consider the underdumped case, $\alpha <\omega .$ Then (\ref{i22}) is
a Hamiltonian of a harmonic oscillator with an angular frequency $\tilde{%
\omega}=\sqrt{\omega ^{2}-{\alpha }^{2}}$. Stationary states of the
corresponding Schr\"{o}dinger equation have the form 
\begin{eqnarray}
&&\Psi _{n}=e^{-iE_{n}t}\psi _{n}(Q)\,,\;E_{n}=\tilde{\omega}\left( n+\frac{1%
}{2}\right) \,,\ \ n=0,1,...,  \notag \\
&&\psi _{n}(Q)=\frac{1}{\sqrt{2^{n}n!}}\left( \frac{\tilde{\omega}}{\pi }%
\right) ^{1/4}e^{-\tilde{\omega}Q^{2}/2}H_{n}\left( Q\sqrt{\tilde{\omega}}%
\right) \,,  \label{i23}
\end{eqnarray}
The mean values of energy (of the operator (\ref{i25})) in such states can
be easily calculated, 
\begin{equation}
\left\langle E\right\rangle _{n}=\left( n+\frac{1}{2}\right) \frac{\omega
^{2}}{\tilde{\omega}}e^{-2\alpha t}\,.  \label{E}
\end{equation}
At each fixed time instant, the energy spectrum is discrete, however, it
decreses with time exactly as in classical theory. The same conclusion was
derived in \cite{Hasse}-\cite{Dekker} where a second-order action obtained
by integrating-multiplier method was taken as a starting point for a
quantization. A quantization of the damped oscillator following Bateman (see
above) meets serious difficulties such as indefinite metric etc., \cite%
{Dekker, Celeghini}.

Overdumped cases, when $\alpha \geq \omega $ correspond to an aperiodic
motion in classical theory \cite{LL1}. Its quantum interpretation is not
clear due to continuous character of the Hamiltonian spectrum.

A nontrivial generalization of equation (\ref{dos}) could be an $n$%
-dimensional damped oscillator\footnote{%
Here, we use matrix notation,{\LARGE \ }$r=\left( r^{i}\right) ,\;\mathbf{a}%
=\left( a_{j}^{i}\right) ,\;\mathbf{\omega }=\left( \omega _{j}^{i}\right)
,\;i,j=1,..,n$.
\par
{}} 
\begin{equation}
\ddot{r}+2\mathbf{a}\dot{r}+\mathbf{\omega }r=0\,,  \label{dos1}
\end{equation}
with the matrixes $\mathbf{a}$\ and $\mathbf{\omega }$\ being constant and
symmetric. Introducing auxiliary variables, $y=\dot{r},$\ $y=\left(
y^{i}\right) $, we reduce (\ref{dos1}) to the following set of first-order
equations: 
\begin{eqnarray}
\dot{x} &=&Ax\mathbf{\,,}\;(x=\left( x^{\alpha }\right) =(r^{i},y^{i}))\,,
\label{i26} \\
A &=&\left( 
\begin{array}{cc}
0 & \mathbf{I} \\ 
-\mathbf{\omega } & -2\mathbf{a}%
\end{array}
\right) \,.  \notag
\end{eqnarray}
This is a set of linear equations with constant coefficients. In this case,
the corresponding quantum Hamiltonian can be constructed, according to (\ref%
{a50}), as follows: 
\begin{equation}
\hat{H}=\frac{1}{2}\hat{x}_{0}B\left( 0\right) \hat{x}_{0}\,,  \label{i27}
\end{equation}
where 
\begin{equation*}
B\left( 0\right) =\left( 
\begin{array}{cc}
\mathbf{\omega } & \mathbf{a} \\ 
\mathbf{a} & \mathbf{I}%
\end{array}
\right) \,,
\end{equation*}
and the operators $\hat{x}_{0}$\ can be divided into the operators of the
coordinates proper $\hat{q}$\ and those of the corresponding momenta $\hat{p}
$,\ with the canonical commutation relations (\ref{cc1}). Further solution
of the quantum problem with quadratic Hamiltonian (\ref{i27}) can follow,
for example, \cite{Lewis,Chernicov}.

\section{Quantization of radiating point-like charge}

Equations of motion for a nonrelativistic particle moving in electric $%
\mathbf{E}$ and magnetic $\mathbf{H}$ fields with account taken of the back
reaction of the radiation emitted by the particle have the form \cite%
{Ginzburg} 
\begin{eqnarray}
&&m\mathbf{\ddot{r}}=\mathbf{F}+\mathbf{f\,,}\;(\mathbf{r}=(x,y,z))\,, 
\notag \\
&&\mathbf{F}=e\mathbf{E}+\frac{e}{c}[\mathbf{\dot{r}}\times \mathbf{H}],\;%
\mathbf{f=}\frac{2e^{2}}{3c^{3}}\mathbf{\dddot{r}}\,.  \label{f4}
\end{eqnarray}
Here $\mathbf{F}$ is the Lorentz force, $\mathbf{f}$ is the force of the
back reaction of the radiation, $e$ is the charge of the particle, and $c$
is the light velocity. Derivatives with respect to time are denoted by dots
above.

These equations are of third order, therefore a trajectory of a charged
particle cannot by uniquely specified only by initial position and velocity
of the particle. It was also pointed out that together with physically
meaningful solutions, equations (\ref{f4}) have a set of nonphysical
solution \cite{Ginzburg}. However, in the case when the back reaction force $%
\mathbf{f}$ is small when compared to the Lorentz one $\mathbf{F},$%
\begin{equation}
\left| \mathbf{f}\right| \ll \left| \mathbf{F}\right| \,,  \label{f4b}
\end{equation}
these equations can be reduced to the second order equations by means of a
reduction of order procedure. Then the above mentioned problem with
nonphysical solutions does not appear. In the reduction procedure, equations
(\ref{f4}) are replaced by second-order equations $\mathbf{\ddot{r}}=\mathbf{%
g}(\mathbf{r},\dot{\mathbf{r}},e)$ such that all the solutions of the latter
equations would be solutions of (\ref{f4}). The last requirement implies a
partial differential equation on the function $\mathbf{g}(\mathbf{r},\dot{%
\mathbf{r}},e)$ having a unique solution with the natural condition $\mathbf{%
g}(\mathbf{r},\dot{\mathbf{r}},0)=0$, see e.g. \cite{Ginzburg,kup}.

Consider, for example, a particular case $\mathbf{E}=0$, $\mathbf{H}=(0,0,H=%
\mathrm{const})$. In such a case, the reduced second-order equations have
the form \cite{kup} 
\begin{eqnarray}
&&\ddot{x}=-\alpha \dot{x}-\beta \dot{y}\,,\;\ddot{y}=\beta \dot{x}-\alpha 
\dot{y}\,,\;\ddot{z}=0\,,  \notag \\
&&\alpha =\frac{\sqrt{6}\sqrt{3+\sqrt{9+64e^{6}H^{2}}}-6}{8e^{2}}\approx 
\frac{2}{3}e^{4}H^{2}\,,\;\beta =\frac{eH\sqrt{6}}{\sqrt{3+\sqrt{%
9+64e^{6}H^{2}}}}\approx eH\,.  \label{f5}
\end{eqnarray}
Here we have set $m=c=1$ for simplicity. Since the evolution along the $z$%
-axis represents the free motion and decouples from the dynamics in the $xy$%
-plane, we restrict our consideration to the first two equations. At $\alpha
=0$, equations (\ref{f5}) are Lorentz equations with an ``effective''\
magnetic field \textbf{$\beta $ }$=(0,0,\beta /e)$. In this case, the
trajectories are concentric circles. If $\alpha \neq 0,$ the particle
spirals at the origin of $xy$-plane. So, it is natural to treat $\alpha $ as
a friction coefficient.

In order to construct an action functional for the non-Lagrangian
second-order equations (\ref{f5}), we introduce new variables as follows: 
\begin{equation*}
p=\dot{x}+\frac{\beta }{2}y\,,\;q=\dot{y}-\frac{\beta }{2}x\,.
\end{equation*}
In the new variables, we have a set of first-order equations, 
\begin{eqnarray}
\dot{x} &=&p-\frac{\beta }{2}y\,,\;\dot{y}=q+\frac{\beta }{2}x\,,  \notag \\
\dot{p} &=&-\frac{\beta }{2}q-\frac{\beta ^{2}}{4}x-\alpha \left( p-\frac{%
\beta }{2}y\right) \,,\;\dot{q}=\frac{\beta }{2}p-\frac{\beta ^{2}}{4}%
y-\alpha \left( q+\frac{\beta }{2}x\right) \,.  \label{f6a}
\end{eqnarray}
According to general formulas (\ref{i14}), (\ref{i15}), and (\ref{i15}), we
construct the following action for the set (\ref{f6a}): 
\begin{eqnarray}
&&S=\frac{1}{2(\alpha ^{2}+\beta ^{2})}\int e^{\alpha t}\left[ a(p\dot{x}-x%
\dot{p}+q\dot{y}-y\dot{q})+\beta (q\dot{x}-x\dot{q}+y\dot{p}-p\dot{y})\right.
\label{f7} \\
&&+\left. c(p\dot{q}-q\dot{p})+d(x\dot{y}-y\dot{x}%
)-e(p^{2}+q^{2})-f(x^{2}+y^{2})-g(px+qy)-j(qx-py)\right] dt\,,  \notag
\end{eqnarray}
where time-dependent functions $a,b,c,d,e,f,g,$ and $j$ are 
\begin{eqnarray*}
&&a=\alpha ^{2}\cos (\beta t)+\beta ^{2}\cosh \left( -\alpha t\right)
,\;b=\alpha ^{2}\sin (\beta t)+\alpha \beta e^{-\alpha t}-\alpha \beta
\,\cos (\beta t), \\
&&c=e^{-\alpha t}\beta -2\alpha \,\sin (\beta t),\;d=-\frac{\beta ^{3}}{2}%
\sinh \left( -\alpha t\right) -\frac{1}{2}\alpha \beta \sin (\beta t)+\alpha
^{2}\beta \left[ \cos (\beta t)-e^{-\alpha t}\right] , \\
&&e=e^{-\alpha t}\beta ^{2}+\alpha ^{2}\cos (\beta t)-\alpha \beta \,\sin
(\beta t),\;g=\alpha \beta ^{2}\cos (\beta t)+\alpha ^{3}\cos (\beta t), \\
&&f=\frac{\beta }{4}\left\{ \beta ^{3}e^{\alpha t}+\beta \alpha ^{2}\cos
(\beta t)+\alpha \lbrack \beta ^{2}+2\alpha ^{2}]\sin (\beta t)\right\}
,\;j=\alpha ^{3}\sin (\beta t)+e^{\alpha t}\beta ^{3}+\alpha ^{2}\beta
\,\cos (\beta t).
\end{eqnarray*}
In the limit of zero friction $\alpha \rightarrow 0$, this action is reduced
to the usual action for a charged particle in a homogeneous magnetic field $%
\beta $.

The set (\ref{f6a}) is linear one with constant coefficients. For such a
case, the corresponding quantum Hamiltonian can be constructed according to (%
\ref{a50}) as 
\begin{equation}
\hat{H}=\frac{1}{2}\left[ \hat{p}_{1}^{2}+\hat{p}_{2}^{2}+\frac{\alpha }{2}%
\left( \hat{p}_{1}\hat{q}_{1}+\hat{q}_{1}\hat{p}_{1}+\hat{p}_{2}\hat{q}_{2}+%
\hat{q}_{2}\hat{p}_{2}\right) +\beta (\hat{p}_{2}\hat{q}_{1}-\hat{p}_{1}\hat{%
q}_{2})+\frac{\beta ^{2}}{4}\left( \hat{q}_{1}^{2}+\hat{q}_{2}^{2}\right) %
\right]   \label{f8}
\end{equation}%
with operators $\hat{q}_{i}$ and $\hat{p}_{j}$ obeying canonical
commutations relations, 
\begin{equation*}
\left[ \hat{q}_{i},\hat{p}_{j}\right] =i\delta _{ij},\;\left[ \hat{q}_{i},%
\hat{q}_{j}\right] =\left[ \hat{p}_{i},\hat{p}_{j}\right] =0,\ \ i,j=1,2~.
\end{equation*}%
By help of a canonical transformation $(\hat{p}_{1},\hat{q}_{1};\hat{p}_{2},%
\hat{q}_{2})\rightarrow (\hat{p},\hat{x};\hat{q},\hat{y})$, where 
\begin{equation*}
\hat{p}=\hat{p}_{1}+\frac{{\alpha }}{2}{\hat{q}}_{1}{,\;\hat{x}=\hat{q}}_{1}{%
,\;\hat{q}=\hat{p}}_{2}+\frac{\alpha }{2}\hat{q}_{2},\;\hat{y}=\hat{q}_{2}{\
,}
\end{equation*}%
we reduce (\ref{f8}) to the form 
\begin{equation}
\hat{H}=\frac{1}{2}\left[ \hat{p}^{2}+\hat{q}^{2}+\beta (\hat{q}\hat{x}-\hat{%
p}\hat{y})+\frac{\beta ^{2}-\alpha ^{2}}{4}\left( \hat{x}^{2}+\hat{y}%
^{2}\right) \right] .  \label{f9}
\end{equation}%
Condition (\ref{f4b}) in the case under consideration implies $\alpha \ll
\beta $, that is why $\alpha ^{2}$ will be omitted in (\ref{f9}) in what
follows.

Consider eigenstates $\Psi $ for two mutually commuting operators $\hat{H}$
and $\hat{L}=\hat{p}\hat{y}-\hat{q}\hat{x},$%
\begin{equation}
\hat{H}\Psi =E\Psi \,,\;\hat{L}\Psi =M\Psi \,.  \label{f10}
\end{equation}
It is convenient to perform the following canonical transformation $(\hat{p},%
\hat{x};\hat{q},\hat{y})\rightarrow (\hat{P},\hat{X};\hat{Q},\hat{Y})$, 
\begin{eqnarray*}
&&\hat{P}=\hat{p}-\frac{\beta }{2}\hat{y}\,,\;\hat{X}=\frac{1}{\beta }\left( 
\hat{q}+\frac{\beta }{2}\hat{x}\right) \,, \\
&&\hat{Q}=\hat{q}-\frac{\beta }{2}\hat{x}\,,\;\hat{Y}=\frac{1}{\beta }\left( 
\hat{p}+\frac{\beta }{2}\hat{y}\,\right) .
\end{eqnarray*}
It is easy to see that 
\begin{equation*}
\hat{H}=\frac{1}{2}\left( \hat{P}^{2}+\beta ^{2}\hat{X}^{2}\right) \,,\;\hat{%
L}=\beta ^{-1}(\hat{H}_{1}-\hat{H})\,,\;\hat{H}_{1}=\frac{1}{2}\left( \hat{Q}%
^{2}+\beta ^{2}\hat{Y}^{2}\right) \,.
\end{equation*}
Operators $\hat{H}$ and $\hat{H}_{1}$ are Hamiltonians of two independent
harmonic oscillators. Then we can divide variables solving (\ref{f10}).
Thus, we obtain solution of the eigenvalue problem (\ref{f10}), 
\begin{equation*}
\Psi =\Psi _{n,l}\left( X,Y\right) =\psi _{n}(X)\psi
_{l}(Y)\,,\;\;E_{n}=\beta \left( n+\frac{1}{2}\right)
\,,\;M_{nl}=l-n\,,\;n,l=0,1,2,...
\end{equation*}
where $\psi _{n}$ and $\psi _{l}$ are eigenstates of the Hamiltonians $\hat{H%
}$ and $\hat{H}_{1}$ respectively (given e.g. by (\ref{i23})). Finally,
stationary states $\Psi (t)$ of the corresponding Schr\"{o}dinger equation
with the Hamiltonian $\hat{H}$\ have the form 
\begin{equation}
\Psi \left( X,Y,t\right) =e^{-iE_{n}t}\Psi _{n,l}\left( X,Y\right) \,.
\label{f11}
\end{equation}

We define the classical energy $E$ of the system under consideration
according to \cite{LL1} as the mechanical energy of the system without
friction, 
\begin{equation*}
E=\frac{1}{2}\left[ p^{2}+q^{2}+\beta (qx-py)+\frac{\beta ^{2}}{4}\left(
x^{2}+y^{2}\right) \right] \,.
\end{equation*}
One can see that the energy depends of time as follows: $E=E_{0}e^{-2\alpha
t}$. An operator $\hat{E}$ that corresponds to the classical quantity $E$
reads: 
\begin{equation*}
\hat{E}=\frac{1}{2}\left[ \hat{P}^{2}+\beta ^{2}\hat{X}^{2}+\alpha \left( 
\hat{X}\hat{Y}-\hat{P}\hat{Q}\right) \right] e^{-2\alpha t}-2\alpha \left( 
\frac{\alpha }{\beta }\right) \left[ \hat{P}\hat{Y}+\hat{Q}\hat{X}\right]
e^{-2\alpha t}+o\left( \frac{\alpha }{\beta }\right) \,.
\end{equation*}
Mean values of this operator in stationary states (\ref{f11}) can be easily
calculated, they are 
\begin{equation*}
\left\langle E\right\rangle _{nl}=\beta \left( n+\frac{1}{2}\right)
e^{-2\alpha t}.
\end{equation*}
Similar to the damped oscillator case considered above, at each fixed time
instant, the energy spectrum is discrete, however, it decreses with time
exactly as in classical theory.

We would like to note that in the work \cite{kup} it was shown that although
an\ action principle for the second-order equations (\ref{f5}) describing a
radiating point-like charge does exist, none of the possible corresponding
Lagrangians in the limit of $\alpha \rightarrow 0$ reduces to the Lagrangian
of a\ particle in a magnetic field modulo a total time derivative. That is,
in the case of a radiating point-like charge a perturbation (in the friction
parameter $\alpha $) of a second-order action does not correspond to a
perturbation of the equations of motion (\ref{f5}). For this reason, we
expect some difficulties with the limit of $\alpha \rightarrow 0$ in the
quantum theory of a radiating point-like charge resulting from quantization
based on an\ action functional in the second-order form (such quantization
for a damped harmonic oscillator was presented in \cite{Dekker}-\cite%
{Khandekar}).

\section{Concluding remarks}

We stress that any nondegenerate set of differential equations written in an
equivalent first-order form can be derived from an action principle. In the
general case, such a set does not provide enough information to fix a class
of quantum theories that, in the classical limit, provide this set of
differential equations for mean values. Therefore, physical considerations
must be used to choose an adeqate quantum theory. In particular, if one
definitely knows that a non-Lagrangian set of equations describes a
dissipative system, which is subjected to a dissipation due to essential
interaction with an environment (reservoir), it is reasonable to consider
the system and the reservoir as two interacting subsystems of a closed
system. Then a quantum description of the dissipative subsystem can be
obtained from a quantum theory of a whole system by averaging over the
reservoir. Such an approach was developed in many articles, see \cite%
{Senitzky}-\cite{Caldeira3}. However, one cannot consider such an approach
as quantization of initial dissipative subsystem, since quantization was
already made for the whole system. In the present article, we consider an
approach where we actually quantize a system with a given set of equations.
It turns out that its ``non-Lagrangian'' behavior is due to a time-dependent
external field. It is a principally different physical situation in
comparison with dissipation of a subsystem. However, quantum theories
obtained from our procedure may be useful to describe some
quantum-mechanical properties of both dissipative systems and
``non-Lagrangian'' systems of other physical nature, like a monopole.

\begin{acknowledgement}
Gitman is grateful to the Brazilian foundations FAPESP and CNPq for
permanent support; Kupriyanov thanks FAPESP for support.
\end{acknowledgement}

\end{document}